\documentclass[12pt]{article}

\def\lsim{\mathrel{\rlap{\lower3pt\hbox{\hskip0pt$\sim$}}
     \raise1pt\hbox{$<$}}}         %less than or approx. symbol
\def\gsim{\mathrel{\rlap{\lower4pt\hbox{\hskip1pt$\sim$}}
     \raise1pt\hbox{$>$}}}         %greater than or approx. symbol

\usepackage{amsmath}
\usepackage{amsfonts}

\begin{document}
\begin{titlepage}

\centerline{\Large \bf No vDVZ Discontinuity in Non-Fierz-Pauli Theories}
\medskip
\medskip

\centerline{Zurab Kakushadze$^\S$$^\dag$$^\P$\footnote{\tt Email: zura@quantigic.com}}

\bigskip

\centerline{\em $^\S$ Quantigic$^\circledR$ Solutions LLC}
\centerline{\em 1127 High Ridge Road \#135, Stamford, CT 06905\,\,\footnote{DISCLAIMER: This address is used by the corresponding author for no
purpose other than to indicate his professional affiliation as is customary in
scientific publications. In particular, the contents of this paper are limited
to Theoretical Physics, have no commercial or other such value,
are not intended as an investment, legal, tax or any other such advice,
and in no way represent views of Quantigic$^\circledR$ Solutions LLC,
the website {\underline{www.quantigic.com}} or any of their other affiliates.}}
\centerline{\em $^\dag$ Department of Physics, University of Connecticut}
\centerline{\em 1 University Place, Stamford, CT 06901}
\centerline{\em $^\P$ Department of Theoretical Physics, A. Razmadze Mathematical Institute}
\centerline{\em I. Javakhishvili Tbilisi State University}
\centerline{\em 6 Tamarashvili Street, 0177 Tbilisi, Georgia}
\medskip
\centerline{(February 27, 2014; revised: August 7, 2014)}

\bigskip
\medskip

\begin{abstract}
{}In theories of massive gravity with Fierz-Pauli mass term at the linearized level, perturbative radially symmetric asymptotic solutions are singular in
the zero mass limit, hence van Dam-Veltman-Zakharov (vDVZ) discontinuity. In this note, in the context of gravitational Higgs mechanism, we argue that in non-Fierz-Pauli theories, which non-perturbatively are unitary, perturbative radially symmetric asymptotic solutions have a smooth
massless limit, hence no vDVZ discontinuity. Perturbative vDVZ discontinuity as an artifact of the Fierz-Pauli mass term becomes evident in the language of constrained gravity, which is the massless limit of gravitational Higgs mechanism.
\end{abstract}
\medskip
PACS: 04.50.-h, 04.50.Gh, 04.50.Kd, 11.25.-w, 11.27.+d
\end{titlepage}

\newpage

\section{Introduction and Summary}

{}A general Lorentz invariant mass term for the graviton $h_{MN}$ in the linearized approximation is of the form
\begin{equation}\label{mass.term}
 -{m^2\over 4} \left[h_{MN}h^{MN} - \beta (h^M_M)^2\right]~,
\end{equation}
where $\beta$ is a dimensionless parameter. Perturbatively, for $\beta\neq 1$ the trace component $h^M_M$ is a propagating ghost, while
it decouples in the Minkowski background for the Fierz-Pauli mass term with $\beta = 1$ \cite{FP}. Gravitational Higgs mechanism \cite{KL, thooft} provides a non-perturbative and fully covariant definition of massive gravity. Non-perturbatively, even for $\beta\neq 1$, the Hamiltonian is bounded from below and the perturbative ghost is an artifact of linearization \cite{Unitarity}.\footnote{\,The full non-perturbative Hamiltonian for the model of \cite{thooft}, which has $\beta=1/2$, in the gravitational Higgs mechanism framework was constructed in \cite{JK} and is expressly positive-definite. Non-perturbative unitarity for general $\beta$ was argued in \cite{Unitarity}.}

{}For $\beta=1$ perturbative radially symmetric asymptotic solutions are singular in the $m\rightarrow 0$ limit: we have the van Dam-Veltman-Zakharov (vDVZ) discontinuity \cite{vDV, Zak} and we must consider non-perturbative solutions \cite{Vain}. In this note, following the method of \cite{ZK.massless}, we argue that for $\beta\neq 1$ perturbative solutions have a smooth massless limit, hence no vDVZ discontinuity. Simply put, the perturbative vDVZ discontinuity is an artifact of the Fierz-Pauli mass term. This becomes particularly evident in the language of constrained gravity, which is the massless limit of gravitational Higgs mechanism \cite{ZK.massless}.

\section{Gravitational Higgs Mechanism}

{}In this section we very briefly review gravitational Higgs mechanism and discuss its massless limit.
We have gravity in $D$ dimensions coupled to scalar fields $\phi^A$,
$A = 0,\dots, D-1$. Coordinate-dependent scalar VEVs break diffeomorphisms spontaneously. Because diffeomorphisms are broken spontaneously, the $D$ scalars $\phi^A$ can be gauge-fixed to their background values, which leaves massive gravity.  The resulting action for gravity is given by
\begin{equation}\label{MG.action}
 S = M_P^{D-2}\int d^Dx \sqrt{-G}\left[ R - \mu^2~V\right]~,
\end{equation}
where $\mu$ has the dimension of mass, and $V$ is a dimensionless ``potential" that makes bulk gravity massive and {\em a priori} is a generic function constructed from the metric $G_{MN}$, antisymmetric tensor density $\epsilon_{M_1\dots M_D}$, and the background metric $E_{MN}$. For our purposes here it will suffice to consider potentials of the form $V = V(X)$, where $X\equiv G^{MN}E_{MN}$. The equations of motion read
\begin{equation}\label{R.eom}
 R_{MN} = \mu^2 \left[V^\prime(X)~E_{MN} +  {{V(X) - X~V^\prime(X)}\over {D-2}}~G_{MN}\right]~,
\end{equation}
with the Bianchi identity
\begin{equation}\label{phi.eom}
 \partial_M\left[\sqrt{-G}V^\prime(X) G^{MN}E_{NS}\right] - {1\over 2}\sqrt{-G}V^\prime(X) G^{MN}\partial_S E_{MN} = 0~,
\end{equation}
which is equivalent to the gauge-fixed equations of motion for the scalars. In (\ref{MG.action}) we have deliberately omitted any source terms. In this note we will focus on the cases with Ricci-flat background metric $E_{MN}$, which implies that $V(D) = 2V^\prime(D)$.

{}In the linearized approximation the r.h.s. of (\ref{R.eom}) corresponds to the graviton mass term (\ref{mass.term}) with
 \begin{eqnarray}\label{m}
 &&m^2\equiv 2\mu^2 V^\prime(D)~,\\
 &&\beta \equiv {1\over 2} - {V^{\prime\prime}(D) \over {V^\prime(D)}}~.\label{beta}
\end{eqnarray}
We have $\beta = 1$ for potentials $V$ with $V^\prime(D) = -2V^{\prime\prime}(D)$. For a linear potential $V(X) = a + X$, we have the model of \cite{thooft} with $\beta = 1/2$. {\em E.g.}, for quadratic potentials $V = a + X + \lambda X^2$ with $\lambda\neq 0$ we can have other values of $\beta$, including $\beta = 1$ for $\lambda = -1/2(D+2)$.

\subsection{Constrained Gravity as the Massless Limit}

{}The massless limit $m \rightarrow 0$ corresponds to taking $\mu\rightarrow 0$. In this limit we obtain not Einstein-Hilbert gravity but {\em constrained} gravity \cite{ZK.massless}. This is because the Bianchi identity (\ref{phi.eom}) survives in the massless limit.
Here $E_{MN}$ is the flat Minkowski metric $\eta_{MN}$
if the coordinates $x^M$ are Minkowski coordinates. However, in general the metric $E_{MN}$ need not be the Minkowski metric. For instance, in spherical coordinates we have
\begin{equation}
 E_{MN}dx^M dx^N = -dt^2 + dr^2 + r^2\gamma_{ab} dx^a dx^b~,
\end{equation}
where $\gamma_{ab}$ is a metric on the unit sphere $S^{d-1}$, $d\equiv D-1$.

{}The fact that we obtain constrained gravity in the massless limit is important. If we take, say, a spherically
symmetric solution in massive gravity and consider the massless limit, it need not coincide with the Schwarzschild
solution of Einstein-Hilbert gravity. Instead, it should coincide with
a spherically symmetric solution in constrained gravity. One way to construct solutions in constrained gravity is to start with known solutions in
Einstein-Hilbert gravity and coordinate-transform them to satisfy the constraint \cite{ZK.massless} (this is similar to \cite{GS}).

\section{Spherically Symmetric Solutions}

{}For spherically symmetric solutions the metric reads ($A, B, C$ are functions of $r$ only):
\begin{equation}
 ds^2 = -A^2 dt^2 + B^2 dr^2 + C^2 \gamma_{ab} dx^a dx^b
\end{equation}
and we have
\begin{eqnarray}
 X =A^{-2} + B^{-2} + (D-2) r^2 C^{-2}~.
\end{eqnarray}
The non-vanishing components of $R_{MN}$
are given by (prime denotes derivative w.r.t. $r$, not to be confused with derivative w.r.t. $X$ as in $V^\prime(X)$):
\begin{eqnarray}
 &&R_{00} = A^2 B^{-2} \left[{A^{\prime\prime}\over A} - {A^\prime B^\prime\over {AB}} + (D-2) {A^\prime C^\prime\over {AC}} \right]~,\\
 &&R_{rr} = -\left\{{A^{\prime\prime}\over A} - {A^\prime B^\prime\over {AB}} + (D-2)
 \left[{C^{\prime\prime}\over C} - {B^\prime C^\prime\over {BC}}\right] \right\}~,\\
 &&R_{ab} = -\gamma_{ab} R_*~,\\
 &&R_*\equiv C^2 B^{-2} \left\{{C^{\prime\prime}\over C} + (D-3) \left({C^\prime\over C}\right)^2 +
 {C^\prime\over C}\left[{A^\prime\over A} - {B^\prime\over B}\right]\right\} - (D-3)~.
\end{eqnarray}
The Bianchi identity (\ref{phi.eom}) reduces to a single equation:
\begin{equation}
 \partial_r\left[AB^{-1}C^{D-2} Q\right] - (D-2)r ABC^{D-4} Q = 0~,
\end{equation}
where $Q\equiv V^\prime(X)$. We will focus on $D=4$ for the remainder of this paper as the generalization to higher $D$ is straightforward.

\subsection{Four-dimensional Massless Solutions}

{}Let us start by analyzing the above equations in the massless case ($m^2 = 0$) in $D=4$. We have the following equations
\begin{eqnarray}
 &&{A^{\prime\prime}\over A} - {A^\prime B^\prime\over {AB}} + 2 {A^\prime C^\prime\over {AC}} = 0~,\\
 &&{A^{\prime\prime}\over A} - {A^\prime B^\prime\over {AB}} + 2
 \left[{C^{\prime\prime}\over C} - {B^\prime C^\prime\over {BC}}\right] = 0~,\\
 &&{C^{\prime\prime}\over C} + \left({C^\prime\over C}\right)^2 +
 {C^\prime\over C}\left[{A^\prime\over A} - {B^\prime\over B}\right] - B^2 C^{-2} = 0~,
\end{eqnarray}
plus the constraint
\begin{equation}\label{const-massless}
 \partial_r\left[AB^{-1}C^2 Q \right] - 2 r AB Q = 0~.
\end{equation}
If it were not for the constraint, we could simply take the Schwarzschild solution:
\begin{eqnarray}
 &&{\overline A} = {\overline B}^{-1} = \sqrt{1-{r_*\over r}}~,\\
 &&{\overline C} = r~.
\end{eqnarray}
However, this solution does not satisfy the constraint.

{}There is a systematic way of finding solutions that satisfy the constraint by transforming known solutions that satisfy unconstrained
Einstein's equations. Thus, we start from a known unconstrained solution given by ${\overline A}, {\overline B}, {\overline C}$,
and transform the radial coordinate $r\rightarrow f(r)$. The resulting metric components are given by
\begin{eqnarray}
 &&A(r) = {\overline A}(f(r))~,\\
 &&B(r) = {\overline B}(f(r)) f^\prime(r)~,\\
 &&C(r) = {\overline C}(f(r))~,
\end{eqnarray}
and they still satisfy the equations of motion. This is because the massless equations of motion possess full reparametrization invariance.
The constraint then produces a second order differential equation for the function $f(r)$. Thus,
starting with the Schwarzschild solution, we can obtain solutions satisfying the constraint by setting $f(r) = C(r)$ in the above
expressions, which gives a differential equation for $C$. We have:
\begin{eqnarray}
 A = \sqrt{1 - r_* / C}~,\\
 B = {C^\prime\over \sqrt{1 - r_* / C}}~,
\end{eqnarray}
and the differential equation for $C$ reads:
\begin{equation}\label{const-massless.11}
 \partial_r\left[A^2 C^2 Q / C^\prime \right] - 2 r Q C^\prime = 0~.
\end{equation}
While (\ref{const-massless.11}) is highly non-linear, we can solve it in two regimes: near the horizon ($C\rightarrow r_*$), and asymptotically ($r \gg r_*$). Here we are interested in asymptotic solutions.

\subsection{Perturbative Asymptotic Solutions}

{}To find perturbative asymptotic solutions to (\ref{const-massless.11}), we set
\begin{equation}\label{lin.c}
 C = r(1 + c)
\end{equation}
and only keep terms linear in $c$. This is equivalent to assuming that $c = \gamma (r_*/r) + {\cal O}(r_*/r)^2$, keeping only the leading terms linear in  $(r_*/r)$ and solving for the coefficient $\gamma$ by requiring that (\ref{const-massless.11}) is satisfied to this approximation. A little straightforward algebra gives
\begin{equation}
 \gamma = {1\over 2}~{V^\prime(4)\over{V^\prime(4) + 2V^{\prime\prime}(4)}}= {1\over 4(1-\beta)}~,
\end{equation}
where we have used (\ref{beta}). So, for $\beta\neq 1$ we have a perturbative asymptotic solution in constrained gravity which is the massless limit of the corresponding perturbative asymptotic solution in massive gravity. This massless perturbative asymptotic solution is valid at distance scales $r\gg r_1 \equiv \gamma r_* = r_*/4(1 - \beta)$. As $\beta\rightarrow 1$, this distance scale $r_1\rightarrow \infty$. This implies that we have the perturbative vDVZ discontinuity for $\beta = 1$, but not for $\beta\not = 1$.

\subsubsection{Non-perturbative Asymptotic Solutions for $\beta=1$}\label{nonpert.massless}

{}The above result shows that for $\beta=1$ the linearized approximation (\ref{lin.c}) breaks down and we must look for non-perturbative asymptotic solutions. We can find a solution via the following Ansatz:
\begin{eqnarray}
 C = r\left[1 + \alpha \left({r_*\over r}\right)^{1\over 2} + \eta~{r_* \over r} + {\cal O}\left({r_*\over r}\right)^{3\over 2}\right]~,
\end{eqnarray}
where $\alpha$ and $\eta$ are numerical coefficients to be determined. This solves (\ref{const-massless.11}) \cite{ZK.massless}:
\begin{eqnarray}
 &&A = 1 - {r_*\over 2r} + {\cal O}\left({r_*\over r}\right)^{3\over 2}~,\\
 &&B = 1 + \sqrt{8\over 39} \left({r_*\over r}\right)^{1\over 2} + {r_* \over 2r} + {\cal O}\left({r_*\over r}\right)^{3\over 2}~,\\
 &&C = r\left[1 + \sqrt{8\over 39} \left({r_*\over r}\right)^{1\over 2} + \eta~{r_* \over r} + {\cal O}\left({r_*\over r}\right)^{3\over 2}\right]~,
\end{eqnarray}
and $\eta$ is an integration constant. This is because we started with the Schwarzschild solution and transformed it via $r \rightarrow C(r)$.
The constraint (\ref{const-massless.11}) is a second order differential equation for $C$, whose solution contains two integration constants. However, because
we drop subleading terms, the resulting equation effectively is only a first order equation for $C$, so we have one integration
constant (and the second integration constant controls the subleading terms). It simply parameterizes the Schwarzschild solution in the transformed coordinate frame.

\subsection{Four-dimensional Massive Solutions}

{}We can derive the above result that there is no perturbative vDVZ discontinuity for $\beta\neq 1$ by directly solving the massive equations of motion in the asymptotic regime. In four dimensions we have:
\begin{eqnarray}
 \label{EOM-mass-A}
 &&A^2 B^{-2} \left[{A^{\prime\prime}\over A} - {A^\prime B^\prime\over {AB}} + 2 {A^\prime C^\prime\over {AC}} \right] =
 \mu^2\left\{A^2~{{XV^\prime(X) - V(X)}\over 2} - V^\prime(X)\right\},\\
 \label{EOM-mass-B}
 &&{A^{\prime\prime}\over A} - {A^\prime B^\prime\over {AB}} + 2
 \left[{C^{\prime\prime}\over C} - {B^\prime C^\prime\over {BC}}\right] = \mu^2\left\{B^2~{{XV^\prime(X) - V(X)}\over 2} - V^\prime(X)\right\},\\
 &&C^2 B^{-2} \left\{{C^{\prime\prime}\over C} +  \left({C^\prime\over C}\right)^2 +
 {C^\prime\over C}\left[{A^\prime\over A} - {B^\prime\over B}\right]\right\} - 1 = \nonumber\\
 &&\,\,\,\,\,\,\,\mu^2\left\{C^2~{{XV^\prime(X) - V(X)}\over 2} - r^2 V^\prime(X)\right\},
 \label{EOM-mass-C}\\
 &&\partial_r\left[AB^{-1}C^2 V^\prime(X)\right] - 2 r AB V^\prime(X) = 0~,
\end{eqnarray}
where the last equations is the Bianchi identity.

\subsubsection{Perturbative Asymptotic Solutions}

{}Let
\begin{eqnarray}
 &&A = 1 + a~,\\
 &&B = 1 + b~,\\
 &&C = r(1 + c)~.
\end{eqnarray}
Here we assume that $a, b, c$ go to zero asymptotically, and in the equations of motion we keep only linear
terms in $a, b, c$. As we will see in a moment, this approximation breaks down for small graviton mass when $\beta = 1$, hence the vDVZ discontinuity.

{}The above four equations of motion in the linearized approximation read:
\begin{eqnarray}
 &&a^{\prime\prime} + {2\over r} a^\prime = m^2 \left[a - \nu z\right]~,\\
 &&a^{\prime\prime} + 2c^{\prime\prime} + {4\over r} c^\prime - {2\over r} b^\prime = m^2 \left[b - \nu z\right]~,\\
 &&c^{\prime\prime} +  {4\over r} c^\prime + {1\over r} \left[a^\prime - b^\prime\right] - {2\over r^2}\left[b - c\right] = m^2 \left[c - \nu z\right]~,\\
 &&a^\prime-b^\prime+2c^\prime - 2\nu z^\prime - {4\over r}\left[b - c\right] = 0~,
\end{eqnarray}
where $z\equiv a + b + 2c$, $\nu\equiv V^{\prime\prime}(4)/V^\prime(4) = 1/2 - \beta$ (see (\ref{beta})), and $m^2 = 2\mu^2 V^\prime(4)$ (see (\ref{m})). From the above four equations we have the following equation for $z$:
\begin{equation}
 2(1-\beta)\left[z^{\prime\prime} + {2\over r}z^\prime\right] = (4\beta - 1)m^2 z~.
\end{equation}
For $\beta =1$ we therefore have $z = 0$, which is simply a manifestation of the fact that perturbatively the trace of the graviton is not a propagating degree of freedom, and
\begin{eqnarray}
 &&a = {\zeta\over r} {\rm e}^{-mr}~,\\
 &&b = {\zeta\over m^2 r^3}\left[1 + m r\right]{\rm e}^{-mr}~,\\
 &&c = -{\zeta\over 2r} {\rm e}^{-mr}-{\zeta\over 2 m^2 r^3}\left[1 + m r\right]{\rm e}^{-mr}~,
\end{eqnarray}
where $\zeta$ is an integration constant. The only way to have a smooth massless limit would be to take $\mu\rightarrow 0$ and
$\zeta\rightarrow 0$ with $|\zeta|/m^2 \equiv r_2^3$ fixed. In this case in the massless limit we would have $a = 0$ and $b = -2c = r_2^3/r^3$. However, the corresponding metric is equivalent to a coordinate-transformed flat metric (in spherical coordinates). So, for $\beta=1$ we have the perturbative vDVZ discontinuity. However, this discontinuity is an artifact of the perturbative approximation, which breaks
down at $r \sim r_2$. Note that $|\zeta|$ is expected to be of order of the Schwarzschild radius $r_*$, so we have $r_2\sim (r_* / m^2)^{1/3}$. This scale goes to infinity when $m$ goes to zero, so one must consider non-perturbative solutions \cite{Vain}.

{}On the other hand, for $\beta\neq 1$ we have no perturbative vDVZ discontinuity. Indeed, for $\beta\neq 1$ and $\beta\neq 1/2$ (so $\nu\ne 0$ and $M\neq m$ -- see below) we have:
\begin{eqnarray}
 &&a = {\zeta\over r} \left[{\rm e}^{-mr} - {1\over 4}{\rm e}^{-Mr}\right],\\
 &&b = {\zeta\over m^2 r^3}\left[\left(1 + m r\right){\rm e}^{-mr} - \left(1 + M r\right){\rm e}^{-Mr}\right] -{3\beta\over 4(1-\beta)}~{\zeta \over r} {\rm e}^{-Mr}~,\\
 &&c = -{\zeta\over 2r} \left[{\rm e}^{-mr} + {1\over 2}{\rm e}^{-Mr}\right]-{\zeta\over 2 m^2 r^3}\left[\left(1 + m r\right){\rm e}^{-mr} -
 \left(1 + Mr\right){\rm e}^{-Mr}\right],
\end{eqnarray}
where $M^2\equiv m^2(4\beta-1)/2(1-\beta)$ is the perturbative mass of the trace $h^M_M$, and $\zeta$ is an integration constant. In the massless limit we have
$a = -r_*/2r$, $b=r_*/2r$ and $c =\gamma r_*/r$, where $r_*\equiv -3\zeta/2$ and $\gamma = 1/4(1-\beta)$, which is the very result we obtained in the beginning of this subsection in constrained gravity.

{}When $\beta = 1/2$, the two masses are degenerate, $M = m$, but the above formulas are still valid. We have $a = -b = -c = \zeta_1\exp(-mr)/r$, where $\zeta_1$ is an integration constant. So for $\beta\neq 1$ we have no perturbative vDVZ discontinuity.

\subsubsection{Non-perturbative Asymptotic Solutions for $\beta = 1$}

{}For $\beta=1$ the linearized approximation breaks down and we must consider non-perturbative massive solutions. In the massless limit they smoothly go to the asymptotic massless solutions we discussed for $\beta=1$ in Subsection \ref{nonpert.massless}. We have:
\begin{eqnarray}
 &&A = 1 - {r_*\over 2r} + {\cal O}\left({r_*\over r}\right)^{3\over 2} + {\cal O}(\mu^2\sqrt{r_* r^3})~,\\
 &&B = 1 + \sqrt{8\over 39} \left({r_*\over r}\right)^{1\over 2} + {r_* \over 2r} + {\cal O}\left({r_*\over r}\right)^{3\over 2} + {\cal O}(\mu^2\sqrt{r_* r^3})~,\\
 &&C = r\left[1 + \sqrt{8\over 39} \left({r_*\over r}\right)^{1\over 2} + \eta~{r_* \over r} + {\cal O}\left({r_*\over r}\right)^{3\over 2}
 + {\cal O}(\mu^2\sqrt{r_* r^3})\right]~,
\end{eqnarray}
where $\eta$ is an integration constant. Note that the expansion in $\mu^2$ is valid at distance scales $r\ll 1/\mu$. As $\mu\rightarrow 0$, we have a smooth massless limit for all $r$.

\subsubsection{Comments}

{}Why is all this useful? If $\beta\neq1$, then asymptotic perturbative computations in cases where the conjugate momenta for the relevant degrees of freedom are small (see below) -- and this includes static solutions -- are valid without invoking the Vainshtein mechanism \cite{Vain}, {\em i.e.}, there is no {\em large} scale -- such as $r_2\sim (r_*/m^2)^{1/3}$ for $\beta = 1$ -- below which the perturbative approximation breaks down. As was argued in \cite{Unitarity}, while for $\beta\neq 1$ the trace $h$ is a ghost, this is a mere artifact of linearization and non-perturbatively the Hamiltonian is bounded from below. Simply put, when relevant conjugate momenta are large (see \cite{Unitarity} for details) -- which is precisely when the ``ghostliness" of $h$ would become problematic -- the perturbative expansion that produces the fake ``ghost" $h$ is invalid in the first place, and non-perturbatively there is no ghost. So {\em a priori} there is no reason to discard $\beta\neq 1$ cases as ``bad". In fact, there is no symmetry that would protect $\beta$ from quantum corrections. In gravitational Higgs mechanism requiring that $\beta = 1$ is nothing but a fine-tuning of the vacuum energy density in the unbroken phase against higher-derivative couplings in the scalar sector \cite{ZK1}, which fine-tuning is unstable against quantum corrections.

%\subsection*{Acknowledgments}

%%%\newpage

\end{document}